\documentclass[conference]{IEEEtran}
\IEEEoverridecommandlockouts
\usepackage{amssymb} 
\usepackage{amsmath} 
\allowdisplaybreaks
\usepackage{hyperref}
\usepackage{nicematrix}
\usepackage{tikz}
\usepackage{xcolor}
\usepackage{booktabs}
\usepackage{enumitem} 
\usepackage{tabularx}
\usepackage{amsmath, amsthm}
\newtheorem{remark}{Remark}

\newtheorem{theorem}{Theorem}
\newtheorem{definition}{Definition}

\newcommand{\linebreakand}{%
  \end{@IEEEauthorhalign}
  \hfill\mbox{}\par
  \mbox{}\hfill\begin{@IEEEauthorhalign}
}
\definecolor{niceblue}{rgb}{0.125, 0.406, 0.852}
\hypersetup{
    colorlinks=true,
    linkcolor=niceblue,
    filecolor=magenta,      
    urlcolor=niceblue,
    citecolor=niceblue}

\newcommand{\tb}[1]{{\color{niceblue}#1}}

\usepackage{float}
\usepackage{bm}
\usepackage{algorithm}
\usepackage{algpseudocode}
\usepackage{booktabs}
\usepackage{multirow}
\newcommand{\btt}[1]{{\fontfamily{lmtt}\selectfont #1}}

\usepackage[table]{xcolor}   
\definecolor{liftred}{RGB}{225, 165, 165}
\definecolor{liftyellow}{RGB}{250, 225, 175}

\definecolor{liftgreen}{RGB}{214,236,222}
\definecolor{solvergray}{RGB}{232,232,232}

\usepackage{amsmath,amssymb,amsfonts}
\usepackage{graphicx}
\usepackage{textcomp}
\usepackage{xcolor}
\def\BibTeX{{\rm B\kern-.05em{\sc i\kern-.025em b}\kern-.08em
    T\kern-.1667em\lower.7ex\hbox{E}\kern-.125emX}}
\usepackage{tikz}
\usetikzlibrary{arrows.meta}

\DeclareRobustCommand{\rightuparrow}{%
  \tikz[baseline=0pt]
    \draw[-{Computer Modern Rightarrow}, line join=miter]
      (0,0.3ex) -- (0.6em,0.3ex) -- (0.6em,1.6ex);%
}

\begin{document}

\title{Low-Rank and Lifted Semidefinite Programming for Mixed-Integer Polynomial Power Grid Optimization\thanks{This work was supported by NSF CAREER (award ECCS-2442592).}}

\author{\IEEEauthorblockN{Samuel Chevalier}
\IEEEauthorblockA{\textit{Electrical and Biomedical Engineering} \\
\textit{University of Vermont}\\
Burlington, VT, USA \\
schevali@uvm.edu}
}

\maketitle

\begin{abstract}
Can a local solver return the guaranteed globally optimal solution to a nonconvex mixed-integer polynomial power grid optimization problem? By mixing low-rank semidefinite programming (SDP) and moment-based lifting in the Lasserre hierarchy, this paper provides anecdotal evidence in the affirmative. Specifically, we iteratively tighten a moment-based relaxation of the AC Optimal Transmission Switching (AC-OTS) problem, including higher order moment-based constraints, rooted in the Lasserre hierarchy, until it is tight (i.e., primal feasible). We then re-solve this problem using a local, low-rank SDP solver (Knitro), plugging the associated dual variables into a bounding Lagrange dual function. By demonstrating primal and dual objective agreement and primal feasibility, we confirm the solution as the exact, global solution. From these initial results, we hypothesize that mixing ($i$) low-rank (i.e., highly scalable) SDP methods with ($ii$) moment-based relaxation tightening could be a scalable alternative to global branch-and-bound-based methods.
\end{abstract}


\section{Introduction}
Branch-and-bound is the dominant approach for globally solving nonconvex optimization problems: given traditional memory constraints, cutting up the problem into a large number of smaller, convex search problems has been the most tractable solution strategy. This approach dominates the commercial integer solver landscape (e.g., Gurobi, Mosek, Knitro, BARON, etc.), but in the worst case, it can be exponentially slow. Because of this general intractability, power grid operators rely on heuristics to solve mixed integer nonlinear programs (MINLPs) (e.g., linearize the problem $\rightarrow$ solve resulting MILP to an acceptable gap $\rightarrow$ project the solution onto the AC-feasible space~\cite{ferc}).

The ARPA-E Grid Optimization competitions~\cite{Holzer2025} sought to address this segmented workflow challenge by fusing standard binary Unit Commitment (UC) on the market clearing side with AC optimal power flow (AC-OPF) on the operations side. Top-performing teams used mixed-integer second order cone (MISOC) relaxations for tuning binaries, cleanup heuristics to restore AC feasibility, and screening methods to identify critical contingencies~\cite{Holzer2025,go-kyri,go3Sam}. Despite extensive participation in the competition, ($i$) the largest and hardest testcases could not be solved by any team, ($ii$) most teams ignored line switching entirely (too hard), and ($iii$) teams were generally happy to find a single, locally feasible solution, much less a globally optimal one, on some of the cases. Using the GO competition as a bellwether for state-of-the-art power grid optimization, real-time optimization of mixed-integer polynomial power grid problems is still an open challenge, with massive potential to improve the efficiency of electricity spot markets, if fully solved~\cite{spot-markets}.  

This paper does not try to ``fully solve" the problem. Instead, we provide initial evidence that a scalable alternative to branch-and-bound might be possible. We focus on the relevant class of nonconvex mixed-integer polynomial problems, since these problems are amenable to moment-based lifting methods from the Lasserre hierarchy. Lasserre~\cite{lass} introduced the moment-SOS hierarchy (now known as the ``Lasserre hierarchy") for solving nonconvex polynomial optimization problems to global optimality.
For a given semidefinite programming (SDP) relaxation, the moment matrix can be multiplied through by a polynomial constraint $f(x) \ge 0$ to yield a further-lifted localizing constraint matrix which must also be positive semidefinite. A finite number of lifts (under mild assumptions) can yield a globally optimal solution~\cite{Nie2014}.

Around 2014, Molzahn and Hiskens~began applying Lasserre hierarchy-based methods to tighten SDP relaxations of the OPF problem~\cite{dan-moment}, eventually expanding to realistic problems with 10k+ constraints~\cite{josz}. Similar lifting methods have efficiently proved the global optimality of local OPF solutions by replacing localizing matrix constraints with determinant-cut approximations~\cite{verify-sdp}. Gopinath et al.~\cite{prove-opt} then showed that mixing relaxation strategies (optimization-based bound tightening of McCormick envelopes + reformulation linearization technique (RLT) cuts + determinant cuts for Lasserre hierarchy lifting) could result in a stronger formulation still, allowing the authors to close the optimality gap to ~0\% in some of the hardest OPF test cases while avoiding SDP solvers.

Despite the success of lifting methods in the power system domain, to the author's knowledge, no existing approach has used them to tackle mixed integer + polynomial optimization problems, which require a higher degree of lifting. Lifting a large problem to a high degree quickly becomes computationally prohibitive: this motivates our use of \textit{low-rank SDP} solution methods in this paper. The low-rank SDP formulation, which was first proposed by Burer \& Monteiro~\cite{Burer2003} in 2003, exploited an interesting observation: if the solution matrix $X$ of a convex SDP is expected to be low-rank, why not just solve directly for the low-rank solution factors? E.g., if $X$ is rank-3, then it can be decomposed as $X=x×x^T+y×y^T+z×z^T$, where $x$, $y$, $z$ are column vectors. The reason for doing this is clear: rather than needlessly solving for $n^2$ matrix variables, we could just solve for $r\times n$ variables, where $r$ is the expected rank of the solution matrix, dramatically increasing the scalability of the approach over a standard SDP method. The history of low-rank SDP is nicely recounted in~\cite{Burer2023TwoDecades}.

Since B\&M factorization was first proposed, strong theoretical performance guarantees (e.g., proof of global optimality for a sufficiently large low-rank search space~\cite{no-spurious}) have bolstered the method. B\&M originally proposed augmented Lagrangian and L-BFGS methods to solve the nonconvex problem~~\cite{Burer2003}; more recent methods have successfully used primal splitting (ADMM) with GPU-accelerated conjugate gradient linear sub-solves~\cite{lorads,culorads} to scale up LR-SDP to unprecedented dimensions (relative to state-of-the-art SDP scalability), solving SDPs with 100+ million constraints and 10+ million in matrix dimension. LR-SDP has only been applied to OPF in a few small-scale studies~\cite{lropf1,lropf2}. 

This paper is built on the following premise: low-rank SDP and moment-based lifting methods are a good match, since they combine the demonstrated scalability of low-rank methods with the promised relaxation tightening of the moment-based lifting machinery. Key contributions are summarized:
\begin{enumerate}
    \item We propose a moment-based lifting strategy, free from localizing matrices, that can be efficiently solved via low-rank SDP.
    \item We construct a Lagrange dual associated with a mixed low-rank SDP + RSOC problem formulation for testing solution optimality.
    \item Most importantly, we demonstrate that an actual mixed-integer polynomial AC-OTS problem can be solved to guaranteed global optimality with a local solver.
\end{enumerate}

The remainder of the paper outlines the guiding mixed-integer polynomial problem (AC-OTS) and proposes a lifting strategy that is amenable to low-rank SDP methods. Test results from a toy test grid are provided.

\section{AC Optimal Transmission Switching}

This section presents a model of a canonical mixed-integer polynomial power grid optimization problem: AC Optimal Transmission Switching (AC-OTS). The model is presented in its full, nonconvex trilinear form, and then it is relaxed into a lifted SDP.

\subsection{AC-OTS Formulation}
In constructing the AC-OTS model, we make three notable modeling choices:
\begin{enumerate}
    \item The problem is explicitly formulated using only voltage and binary variables (i.e., no auxiliary line flow or generation variables). The big-M reformulation is used to derive implied cuts, but it doesn't explicitly model the problem.
    \item Line flows are limited by current magnitude.
    \item We use rotated second order cones to model generator quadratic costs.
\end{enumerate}

To construct the AC-OTS model, we consider a power grid model with generator set $\mathcal G$, $|{\mathcal G}|=n_g$,  bus set $\mathcal B$, $|{\mathcal B}|=n_b$, line set ${\mathcal L}$, $|{\mathcal L}|=n_l$, rectangular voltage vectors $v_d\in {\mathbb R}^{n_b}$, $v_q\in {\mathbb R}^{n_b}$, and line binary vector $\beta\in {\mathbb R}^{n_l}$. We define power flows using rectangular voltage coordinates ($v_{d,i}, v_{q,i}$) and complex tap ratio ($\tau_{ij} e^{j \theta_{ij}}$). For compact representation, we define the following tri-linear products, which represent the product of network voltages with associated line binary variables: 
\begin{subequations}
\begin{align}
w_{ij} & \triangleq \beta_{ij}\left(v_{d,i}v_{d,j}+v_{q,i}v_{q,j}\right)&&\!\!\!\!\!\rightarrow \beta_{ij}V_{i}V_{j}\cos(\theta_{i}-\theta_{j})\\
u_{ij} & \triangleq \beta_{ij}\left(v_{q,i}v_{d,j}-v_{d,i}v_{q,j}\right)&&\!\!\!\!\!\rightarrow \beta_{ij}V_{i}V_{j}\sin\left(\theta_{i}-\theta_{j}\right)\\
z_{ij} & \triangleq \beta_{ij}\left(v^{2}_{d,i}+v^{2}_{q,i}\right)&&\!\!\!\!\!\rightarrow \beta_{ij}V^{2}_{i}\\
{z}_{ii} & \triangleq \left(v^{2}_{d,i}+v^{2}_{q,i}\right)&&\!\!\!\!\!\rightarrow V^{2}_{i}.
\end{align}
\end{subequations}
Using the complex tap ratios $\tau_{ij}e^{j\theta_{ij}}=t_{d,ij}+jt_{q,ij}$, we define the transformer-modified line parameters via
\begin{align*}
g_{d,ij} & \triangleq g_{ij}t_{d,ij}/\tau^{2}_{ij},\quad g_{q,ij}\triangleq g_{ij}t_{q,ij}/\tau^{2}_{ij}\\
b_{d,ij} & \triangleq b_{ij}t_{d,ij}/\tau^{2}_{ij},\quad b_{q,ij}\triangleq b_{ij}t_{q,ij}/\tau^{2}_{ij}.
\end{align*}
The ``from" $(ij)$ and ``to" $(ji)$ side line flows are now compactly given by
\begin{align}
p_{ij} & \!\triangleq\!\frac{g_{ij}\!+\!g_{i}}{\tau^{2}_{ij}}z_{ij}+(b_{q,ij}-g_{d,ij})w_{ij}-(b_{d,ij}+g_{q,ij})u_{ij}\\
p_{ji} & \!\triangleq\!(g_{ij}\!+\!g_{j})z_{ji}-(g_{d,ij}+b_{q,ij})w_{ij}\!-\!(g_{q,ij}\!-\!b_{d,ij})u_{ij}\\
q_{ij} & \!\triangleq\!-\frac{b_{ij}\!+\!b_{i}}{\tau^{2}_{ij}}z_{ij}\!+\!(b_{d,ij}+g_{q,ij})w_{ij}+(b_{q,ij}\!-\!g_{d,ij})u_{ij}\\
q_{ji} & \!\triangleq\!-(b_{ij}\!+\!b_{j})z_{ji}\!-\!(g_{q,ij}\!-\!b_{d,ij})w_{ij}\!+\!(g_{d,ij}\!+\!b_{q,ij})u_{ij}
\end{align}
where all flows are explicit functions of the voltage and binary variables, rather than new variables. In the appendix, we derive expressions for squared current magnitude ($I^2_{ij}$ and $I^2_{ji}$), which we use to constrain line flows. We constrain current magnitude, rather than apparent power, because $I^2_{ij}$ is a second order function of voltage (bilinear), while $s^2_{ij}$ is a fourth order function of voltage (containing quartic monomials). 

Using the power injection equations, we define the generation injections into bus $i$, as in~\cite{dan-moment}:
\begin{subequations}
\begin{align}
p_{g,i}(v_{d},v_{q},\beta)&\triangleq p_{d,i}+g_{s,i}z_{ii}+\sum_{(ij)\in\mathcal{L}_{i}}p_{ij},\quad\forall i\in {\mathcal B}\\
q_{g,i}(v_{d},v_{q},\beta)&\triangleq q_{d,i}-b_{s,i}z_{ii}+\sum_{(ij)\in\mathcal{L}_{i}}q_{ij}, \quad\forall i\in {\mathcal B}
\end{align}
\end{subequations}
where $\mathcal{L}_{i}$ is the set of lines tied to bus $i$. We can then express the quadratic cost associated with the generator at bus $i$ via
\begin{align}
f_{c,i}(v_{d},v_{q},\beta)\triangleq c_{2,i}p^{2}_{g,i}+c_{1,i}p_{g,i}+c_{0,i}.
\end{align}
Finally, we state the full AC-OTS problem as
\begin{subequations}\label{eq: ac-ots}
\begin{align}\min_{\beta,v_{d},v_{q}}\quad & \sum_{i\in\mathcal{G}}f_{c,i}(v_{d},v_{q},\beta)\\
{\rm s.t.}\;\quad & \underline{p}_{g,i}\le p_{g,i}(v_{d},v_{q},\beta)\le\overline{p}_{g,i}, &\forall i \in {\mathcal B}\\
 & \underline{q}_{g,i}\le q_{g,i}(v_{d},v_{q},\beta)\le\overline{q}_{g,i}, &\forall i \in {\mathcal B}\\
 & \underline{v}^{2}_{i}\le v^{2}_{d,i}+v^{2}_{q,i}\le\overline{v}^{2}_{i}, &\forall i \in {\mathcal B}\\
 & I^{2}_{ij}(v_{d},v_{q},\beta)\le\overline{I}^{2}_{ij}, &\forall (ij) \in {\mathcal L}\label{eq:iij}\\
 & I^{2}_{ji}(v_{d},v_{q},\beta)\le\overline{I}^{2}_{ji}, &\forall (ij) \in {\mathcal L}\label{eq:iji}\\
 & \beta\in\{0,1\}^{n_l}.\label{eq: bin-con}
\end{align}
\end{subequations}
This formulation assumes one generator per bus, and non-generator buses have $\underline{p}_{g,i}=\overline{p}_{g,i}=\underline{q}_{g,i}=\overline{q}_{g,i}=0$. This problem can be canonicalized by forming the vector $x$ and constructing the outer product matrix $X$ as
\begin{align}\label{eq: Xbig}
x=\left[\begin{array}{c}
v_{d}\\
\tilde{v}_{q}\\
\beta
\end{array}\right],\quad X=\underbrace{\left[\begin{array}{c}
1\\
x\\
x\otimes_{s}x
\end{array}\right]}_{\nu}\left[\begin{array}{c}
1\\
x\\
x\otimes_{s}x
\end{array}\right]^{T},
\end{align}
where the reference voltage ${v}_{q,1}=0$ has been removed to form $\tilde{v}_{q}$, and $\otimes_{s}$ denotes the ``symmetric" Kronecker product (it generates the vector of unique degree-2 monomials).

\begin{definition}[Monomial basis vector]
The vector of monomials used to form the lifted matrix $X$ via outer product in \eqref{eq: Xbig} is called the monomial basis vector. It is denoted $\nu$.
\end{definition}

We state the AC-OTS canonicalization\footnote{For real, generally non-symmetric matrices $A,B$, the Frobenius inner product is $\left\langle A,B\right\rangle={\rm tr}(A^TB)=\sum_{ij} A_{ij}B_{ij}$.} via 
\begin{subequations}\label{eq: acots-canon}
\begin{align}\min_{x,t}\quad & \left\langle C,X\right\rangle +t\\
{\rm s.t.}\,\quad & X \text{ from } \eqref{eq: Xbig}\label{eq: xbig-eq}\\
 & \left\langle A_{i},X\right\rangle =0,&&\forall i\in\mathcal{E}\\
 & \left\langle B_{i},X\right\rangle \ge0,&&\forall i\in\mathcal{I}\\
 & 2\cdot0.5\cdot t\ge\sum_{i\in\mathcal{G}}\left\langle K_{i},X\right\rangle ^{2}\label{eq:rsoc}
\end{align}
\end{subequations}
where $\mathcal{E}$ and $\mathcal{I}$ are the equality and inequality constraint sets implied by \eqref{eq: ac-ots} (matrices $A_i$ and $B_i$ are not stated explicitly), and $C$ is a cost matrix. We replace the binary membership constraint \eqref{eq: bin-con} with the nonconvex equality constraint $\beta_{ij} = \beta^2_{ij}$. Finally, \eqref{eq:rsoc} is the rotated second order cone (RSOC) cut that captures the quadratic generation cost (i.e., $\langle K_{i},X\rangle$ is the generation at bus $i$ scaled by $\sqrt c_{2,i}$). We use an explicit RSOC epigraph here because this constraint is both convex and tight, and we don't generally lift this constraint. Thus, leaving it out of the moment matrix is advantageous. To benchmark our results, we use Gurobi to solve a big-M reformulated version of this problem.

\subsection{AC-OTS Relaxation}
We form an AC-OTS relaxation by ($i$) dropping the nonconvex equality constraint \eqref{eq: xbig-eq} and replacing it with $X \succeq 0$, ($ii$) adding $X_{11}=1$ as an explicit constraint (via matrix $A_0$), ($iii$) adding binary idempotency constraints ($\beta_{ij}=\beta_{ij}^2$), ($iv$) adding the variable linking constraints (e.g., $x_1*x_1=x_1^2$) which naturally arise based on the structure of \eqref{eq: Xbig}, since elements of $xx^T$ and $x\otimes_{s}x$ correspond to the same monomials, even though their lifted moment surrogates are completely different variables. The idempotency and linking constraints are both added to the set $\mathcal V$, imposed in \eqref{eq: V}.

\subsubsection*{Big-M cuts} To further tighten the relaxation (and to add cuts to ``tighten around'' in later sections), we also add explicit big-M cuts. In the lifted moment matrix $X$, we have simultaneous access to both $p_{ij}(v_d,v_q)$ (i.e., the flow on a line, regardless of its binary status), $p_{ij}(v_d,v_q,\beta)$ (i.e., the line flow scaled by a binary), and $\beta$ (i.e., the binary itself). Using these three variables, we may construct the following big-M cuts:
\begin{subequations}\label{eq: bigm2}
\begin{align}
-\overline{M}^p_{ij}(1-\beta_{ij})+p_{ij} & \le p_{ij}(\beta)\le p_{ij}-\underline{M}^p_{ij}(1-\beta_{ij})\\
\underline{M}^p_{ij}\beta_{ij} & \le p_{ij}(\beta)\le\overline{M}^p_{ij}\beta_{ij}
\end{align}
\end{subequations}
where $p_{ij}$ is an assumed function of voltage, and $p_{ij}(\beta)$ is a function of both voltage and the corresponding line binary variable; $p_{ij}$ and $p_{ij}(\beta)$ both represent functions, not variables. This big-M tightening scheme is not necessary for constructing the relaxation; however, the author has seen that using these cuts helps the optimizer tighten the relaxed binary variables $\beta_{ij}$ to their true 1/0 values. The big-M tightening routine we employ in this paper is described in the appendix.

We collect all big-M cuts into the set $\mathcal M$ (all index sets are summarized in Table \ref{tab: sets}). The full relaxation is then given by
\begin{subequations}\label{eq: acots-relaxed}
\begin{align}
\min_{X\succeq0}\quad & \left\langle C,X\right\rangle +t\\
{\rm s.t.}\,\quad & \left\langle A_{0},X\right\rangle =1\\
 & \left\langle A_{i},X\right\rangle =0, &  & \forall i\in\mathcal{E}\cup\mathcal{V}\label{eq: V}\\
 & \left\langle B_{i},X\right\rangle \ge0, &  & \forall i\in\mathcal{I}\cup\mathcal{M}\\
 & 2\cdot0.5\cdot t\ge\sum_{i\in\mathcal{G}}\left\langle K_{i},X\right\rangle ^{2}.\label{eq:rsoc2}
\end{align}
\end{subequations}
This relaxed model does not include the full set of implied binary McCormick cuts (i.e., ones that emerge by constructing the four McCormick cuts from any two binary constraints $0 \le \beta_{ij} \le 1$ and $0 \le \beta_{kl} \le 1$) for all binary combinations; selective cuts from this larger set will be added in the following section.

\begin{table}[t]
  \centering
  \caption{}\label{tab: sets}
  \label{tab:index-sets}
  \begin{tabular}{@{}llp{0.55\linewidth}@{}}
    \toprule
    \textbf{Set} & \textbf{Description of set indices}\\
    \midrule
    $\mathcal{E}$ & Original AC-OTS equality constraints \\
    $\mathcal{I}$ & Original AC-OTS inequality constraints \\
    $\mathcal{V}$ & Equality constraints that link lifted variables (binary/continuous)\\
    $\mathcal{M}$ & Big-M/McCormick inequalities \\
    $\mathcal{R}$ & RLT cuts, where $\mathcal{R}= \mathcal{R}_e \cup \mathcal{R}_n$ ($\mathcal{R}_e$: equality, $\mathcal{R}_n$: inequality) \\
    \bottomrule
  \end{tabular}
\end{table}

\section{Low-Rank and Lifted SDP}
In this section, we lift the relaxation via higher order moment matrices, and we propose a low-rank SDP solution procedure.

\subsection{Moment-Based Lifting}
The solution to the relaxed SDP \eqref{eq: acots-relaxed} is generally loose (see results section). To tighten this relaxation, we may ``lift" the problem using one of two tightening mechanisms:
\begin{enumerate}
    \item \textbf{RLT cuts:} by taking the product of constraints, or by taking the product of a constraint with a monomial, we can generate new valid, lifted equalities or inequalities.
    \item \textbf{Localizing matrices:} by taking the product of a nonnegative polynomial constraint and a PSD moment matrix, we can generate a lifted moment matrix which is PSD.
\end{enumerate}
While localizing matrices impose strictly stronger conditions than the corresponding RLT cuts, they are computationally expensive to implement, since they require the addition of new PSD constraints (i.e., $M_i\triangleq (ax_i+b)\cdot(X)\succeq 0$). In this paper, we target the use of low-rank SDP methods (see next section), where matrices are PSD \textit{by construction}. Therefore, we have no direct mechanism for enforcing $M_i\succeq 0$, except through construction, with the addition of a massive number of new variables and linking constraints. For example: 
\begin{align}\label{eq: localize}
\underbrace{M_{i}\succeq0}_{\text{typical SDP constraint}}\quad\leftrightarrow\quad\underbrace{M_{i}=\sum^r_{j=1}y_{j}y^{T}_{j}}_{\text{low-rank SDP constraint}},
\end{align}
where $y_j$ are new variable vectors. In this paper, we avoid localizing matrices entirely, since the endless addition of new variables, constraints, and sources of nonconvexity do not seem to be worth the marginal tightening they provide. Future work will explore linearized determinant cuts~\cite{det-cuts,lin-acopf} or RSOC cuts~\cite{mixed-sdp-rsoc} to achieve localizing matrix tightening. 

We therefore lift using RLT cuts exclusively. In this paper, we consider a menu of three RLT cut options:
\begin{enumerate}[itemsep=0.5em]
    \item Product of equality constraint $\left\langle A_{i},X\right\rangle=0$ and a variable $x_j$. Result: $x_j\cdot \left\langle A_{i},X\right\rangle=0$.
    \item Product of inequality constraint $\left\langle B_{i},X\right\rangle\ge0$ and nonnegative variable $x_j\ge 0$. Result: $x_j\cdot \left\langle B_{i},X\right\rangle\ge0$.
    \item Product of inequality constraint $\left\langle B_{i},X\right\rangle\ge0$ and squared variable $x_j^2$. Result: $x_j^2\cdot\left\langle B_{i},X\right\rangle\ge0$.
\end{enumerate}
To organize the lifting, we construct ``tuples of tuples" to represent monomials, where each ``inner tuple" represents the power that each variable is lifted to (0 powers are excluded):
\begin{align}
x^{2}_{1}x_{3}\;\leftrightarrow\;((x_{1},2),(x_{3},1)).
\end{align}
We then define a matrix of tuples $T(X)$ that maps lifted variables to their corresponding monomials. Applied to matrix $X$ in \eqref{eq: Xbig}, we have
\setlength{\arraycolsep}{2pt}
\begin{align}\label{eq: TX}
T(X)={\footnotesize \left[\!\begin{array}{cccc}
((0,0)) & ((x_{1},1)) & ((x_{2},1)) & \!\!\cdots\\
((x_{1},1)) & ((x_{1},2)) & ((x_{1},1),(x_{2},1))\\
((x_{2},1)) & ((x_{1},1),(x_{2},1)) & ((x_{2},2))\\
\vdots &  &  & \!\!\ddots
\end{array}\!\right]}.
\end{align}
\setlength{\arraycolsep}{5pt}When we construct an RLT cut, we instantiate a new lifted moment submatrix with an updated tuple submatrix. For example, by forming the RLT cut
\begin{align}
x_{1}\left\langle A_{i},X\right\rangle =\left\langle A_{i},x_{1}X\right\rangle =0,
\end{align}
matrix $A_i$ is now being multiplied by a lifted moment submatrix mapping to monomials given by
\setlength{\arraycolsep}{2pt}
\begin{align*}T({\small x_{1}X})\!=\!{\footnotesize \left[\!\!\begin{array}{cccc}
(\tb{(x_{1},1)}) & ((x_{1},\tb{2})) & (\tb{(x_{1},1)},(x_{2},1)) & \!\!\cdots\\
((x_{1},\tb{2})) & ((x_{1},\tb{3})) & ((x_{1},\tb{2}),(x_{2},1))\\
(\tb{(x_{1},1)},(x_{2},1)) & ((x_{1},\tb{2}),(x_{2},1)) & (\tb{(x_{1},1)},(x_{2},2))\\
\vdots &  &  & \!\!\ddots
\end{array}\!\right]}
\end{align*}
\setlength{\arraycolsep}{5pt}where blue terms highlight updates from the original moment matrix $X$ of \eqref{eq: TX}.

\begin{definition}[Lifted moment submatrix] We refer to $x_{i}X$ as a lifted moment submatrix, since it exists as a hypothetical submatrix within some full lifted moment matrix $\tilde X$.

\end{definition}

Although the entire lifted moment submatrix $x_{1}X$ is generated by the RLT cut, the matrix $A_i$ is generally hyper-sparse, so the vast majority of the terms in $x_{1}X$ will not be used. There is a small critical subset, however, that we need access to, so we must choose a monomial basis vector expansion strategy that directly provides, or internally generates, those terms. To avoid unnecessary terms in the monomial basis vector, we identify the smallest necessary expansion of the monomial basis vector, with terms of the lowest possible order, that provide the monomial terms we need access to. We illustrate the intuition of this approach with two small examples that use the following basis: 
\begin{align}
\nu=\left[\begin{array}{c}
1\\
x_{1}\\
x_{2}
\end{array}\right],\quad X=\left[\begin{array}{ccc}
1 & x_{1} & x_{2}\\
\cdot & x^{2}_{1} & x_{1}x_{2}\\
\cdot & \cdot & x^{2}_{2}
\end{array}\right]\label{eq: X-example}
\end{align}
\begin{itemize}
\item \textit{Example 1: no basis expansion required.} Assume we lift an equality constraint $x_1 + x_2 -1 = 0$ by the basis variable $x_1$. The original symmetric constraint matrix is
\[
A_{l}=\left[\begin{array}{ccc}
-1 & 0.5 & 0.5\\
\cdot & 0 & 0\\
\cdot & \cdot & 0
\end{array}\right].
\]
The new, lifted constraint is $x^2_1 + x_1x_2 -x_1 = 0$. In this case, we already have access to all of these lifted variables in the original moment matrix $X$. Thus, we do not need to add any monomials to the monomial basis vector. The updated constraint matrix ${\tilde A}_l$ is given by 
\[
\tilde{A}_{l}=\left[\begin{array}{ccc}
0 & -0.5 & 0\\
\cdot & 1 & 0.5\\
\cdot & \cdot & 0
\end{array}\right],
\]
and the moment matrix remains unchanged.

\item \textit{Example 2: optimal basis expansion.}
In this example, we have two nonlinear constraints that we want to lift by $x_2$ and $x_1$, respectively:
\begin{align*}
x_{2}[x^{2}_{1}-x_{2}=0] & \quad\rightarrow\quad x_{2}x^{2}_{1}-x^{2}_{2}=0\\
x_{1}[x_{1}-x^{2}_{2}=0] & \quad\rightarrow\quad x^{2}_{1}-x_{1}x^{2}_{2}=0.
\end{align*}
We need access to two new monomials, which are not contained in \eqref{eq: X-example}. A non-optimal decision would be to expand the monomial basis by $x_1^2$ and $x_2^2$, adding \textit{two} new terms. However, if we expand the monomial basis by a \textit{single} term, $x_1x_2$, we generate all needed monomials:
\[
{\tilde \nu}=\left[\begin{array}{c}
1\\
x_{1}\\
x_{2}\\
x_{1}x_{2}
\end{array}\right],\;\tilde{X}=\left[\begin{array}{cccc}
1 & x_{1} & x_{2} & x_{1}x_{2}\\
\cdot & x^{2}_{1} & x_{1}x_{2} & \tb{x^{2}_{1}x_{2}}\\
\cdot & \cdot & x^{2}_{2} & \tb{x_{1}x^{2}_{2}}\\
 &  &  & x^{2}_{1}x^{2}_{2}
\end{array}\right].
\]
Expanding the basis by a single term is the better choice, since it generates fewer new decision variables.
\end{itemize}
In the second example, the term $x_1x_2$ shows up twice in the upper triangle of matrix $\tilde X$. This linking must be accounted for through equality constraints, as we now describe.

\subsubsection*{Identification of linking constraints} Due to binary idempotency, a binary is equal to itself raised to any positive integer power $\ge 1$: $\beta_{ij} = \beta^2_{ij} = \beta^3_{ij} = $ etc. To exploit this, we define a filtering function that is applied to each variable factor in a monomial tuple. Given the set $\mathcal Z$ of indices associated with binary variables, the function \textit{demotes} any binary variable power higher than 1 back to exactly 1:
\begin{align}\label{eq: bin-filt}
\text{binary exponent filter:} \; f_b((x_{i},j))=\begin{cases}
(x_{i},1), & i\in\mathcal{Z}\\
(x_{i},j), & i\notin\mathcal{Z}.
\end{cases}
\end{align}
We do this so that we can identify equivalent entries in the tuples matrix and equality-constrain them equal. To do so, we define a tuple linking function: this function searches across the lifted tuple matrix for equivalent entries and generates equality constraint linking matrices (equivalencies implied by matrix symmetry are skipped). For example, the (blue) blocks in \eqref{eq: links} contain $\tfrac{1}{2}(n^2+n)$ equivalent monomials:
\begin{align}\label{eq: links}
X=\left[{\small\begin{array}{ccc}
1 & x^{T} & \tb{(x\otimes_{s}x)^{T}}\\
x & \tb{xx^{T}} & \cdot\\
\tb{x\otimes_{s}x} & \cdot & \cdot
\end{array}}\right].
\end{align}
The tuple linking function \eqref{eq: linkfunc} returns symmetric equality constraint matrices $A_i$ for each found equivalency. Before applying this matrix, we first pass the full lifted tuple matrix through the binary exponent filter $T_b=f_b(T)$. Next, we apply the linking function to generate the set of linking constraints:
\begin{align}\label{eq: linkfunc}
f_{l}(T_{b})=\{A_{ij,kl}\;|\; T_{b,ij}=T_{b,kl},\;kl\ne ji\}.
\end{align}
Equality constraints are implemented via $\left\langle A_{ij,kl},X\right\rangle=0$. We now provide two small examples of the structure of these linking matrices (for a $3 \times 3$ matrix, where dots indicate 0's):
\begin{itemize}
\item \textit{Linking example 1.} If entries $X_{1,2}$ and $X_{2,3}$ are equal,
\begin{align}
A_{12,23}=\left[\begin{array}{ccc}
. & 0.5 & .\\
0.5 & . & -0.5\\
. & -0.5 & .
\end{array}\right].
\end{align}
\item \textit{Linking example 2.} If entries $X_{2,2}$ and $X_{2,3}$ are equal,
\begin{align}
A_{22,23}=\left[\begin{array}{ccc}
. & . & .\\
. & 1 & -0.5\\
. & -0.5 & .
\end{array}\right].
\end{align}
\end{itemize}

Linking constraints help tighten the underlying relaxation, and their associated dual variables are used in the Lagrange dual to construct a certified lower bound. Rather than equality-constrain moment matrix variables, it may feel tempting to instead have matrix entries share the same underlying variable, thus avoiding the need for equality links altogether. However, this is a nontrivial process in the context of low-rank SDP, where the moment matrix variables are shared across multiple low-rank basis vectors. Furthermore, constructing the Lagrange dual from a formulation that shares variables across a matrix quickly becomes cumbersome, with additional equality constraints needed to link shared variables with their low-rank basis terms (similar in spirit to \eqref{eq: localize}), thus undoing any computational efficiency gained by invoking shared variables. 

The overarching RLT-based lifting routine is summarized in the following algorithm, whose style is inspired by~\cite{boxalg}.
\begin{figure}[!t]
\centering
\renewcommand{\arraystretch}{1.15}
\begin{tabularx}{\columnwidth}{|@{\hspace{5pt}}l@{\hspace{5pt}}X@{\hspace{5pt}}|}
\hline
\multicolumn{2}{|c|}{\rule{0pt}{12pt}\textbf{SDP lifting with RLT cuts}}\\[3pt]
\hline
\rule{0pt}{12pt}%
Given: &
The lifted moment matrix $X$\\[4pt]
Step 1: &
\textbf{Choose cuts.} Based on binary fuzziness, constraint tightness (via dual variable magnitude), voltage submatrix rank, or other heuristics, choose a set of RLT cuts to apply $\rightarrow$ update sets ${\mathcal R}_e$ and ${\mathcal R}_n$\\
Step 2: &
\textbf{Lift.} For each RLT cut, generate the associated lifted moment submatrices (e.g., $x_iX$, $x^2_jX$, etc.)\\
Step 3: &
\textbf{Collect missing monomials.} Based on the sparsity pattern of the RLT cut matrices, identify and collect monomial terms from the lifted moment submatrices that are, simultaneously, actually needed in the lift \textit{and} missing from $T(X)$\\
%
Step 4: &
\textbf{Expand monomial basis vector.} Determine the smallest necessary expansion of the monomial basis vector $\nu$ of \eqref{eq: Xbig}, with terms of the lowest possible order, to ensure the RLT cuts can be constructed\\
Step 5: & \textbf{Rebuild coefficient matrices.} Using the expanded monomial basis vector ${\tilde \nu}$ and the associated lifted moment matrix ${\tilde X}$, rebuild all optimization matrices $C$, $A_i$, $B_i$, $K_i$ to ensure correct Frobenius inner product mappings: $\langle A_i,{\tilde X}\rangle=0$, $\langle B_i,{\tilde X}\rangle\ge0$, etc.\\
Step 6: & \textbf{Add linking constraints.} Apply the binary exponent filter \eqref{eq: bin-filt} to the lifted tuple matrix $T(\tilde X)$, and then identify all new equality constraint linking matrices via \eqref{eq: linkfunc} $\rightarrow$ update set $\mathcal V$\\
Step 7: & \textbf{Optionally, drop old, inactive constraints.}\\
Step 8: & \textbf{Solve} newly-lifted SDP and check convergence criteria $\rightarrow$ return to step 1 \rightuparrow{} or terminate \qed\\
\hline
\end{tabularx}
\vspace{-6pt}
\caption{}
\label{fig:sdp-lifting}
\end{figure}

\subsection{Low-rank SDP}
This subsection is devoted to step 8 in the preceding algorithm, which solves the lifted SDP. A conventional conic SDP solver can be used, but this paper focuses on exploiting the potential scalability of low-rank SDP methods. In the low-rank context, the problem we pose in step 8 can be stated as 
\begin{subequations}\label{eq: acots-lr}
\begin{align}
c_p=\min_{x,t} \; & \left\langle C,X\right\rangle +t\\
{\rm s.t.}\;  & \lambda_0\!: \left\langle A_{0},X\right\rangle =1\\
 & \lambda_i\!:\,\left\langle A_{i},X\right\rangle =0, &  & \!\!\!\!\!\!\!\!\!\!\!\!\!\!\!\!\forall i\in\mathcal{E}\cup\mathcal{V}\cup{\mathcal R}_e\label{eq: Vlr}\\
 & \mu_i\!:\,\left\langle B_{i},X\right\rangle \ge0, &  & \!\!\!\!\!\!\!\!\!\!\!\!\!\!\!\!\forall i\in\mathcal{I}\cup\mathcal{M}\cup{\mathcal R}_n\\
 & s\!:\;\;t\ge\sum\nolimits_{i\in\mathcal{G}}\left\langle K_{i},X\right\rangle ^{2}\label{eq:rsoc2lr}\\
 & \quad \;\; X=\sum\nolimits_{i=1}^{r} x_i x_i^T
\end{align}
\end{subequations}
which is similar in structure to the original relaxation \eqref{eq: acots-relaxed}, but with $(i)$ the low-rank factorization directly embedded, $(ii)$ the RLT cut sets (${\mathcal R}_e$, ${\mathcal R}_n$) explicitly added, and $(iii)$ the dual variables explicitly noted ($s$ is a tuple\footnote{The tuple $s=(s_1,s_2,y)$ holds the dual variables associated with the RSOC constraint. If $(t,0.5,[K_1,K_2,...,K_n])\in {\mathcal K}$ and $s\in {\mathcal K}^*$, where ${\mathcal K}$, ${\mathcal K}$ are the primal, dual RSOCs, then $s_{1}t+s_{2}\cdot0.5+\sum_{i}y_{i}\left\langle K_{i},X\right\rangle \ge0$.} of dual RSOC variables~\cite{mosek_modeling_cookbook}). Since this problem is nonconvex, there is generally no guarantee that a local solution is indeed a global solution. After solving the primal problem \eqref{eq: acots-lr} with a low-rank SDP solver, our strategy is to pass the corresponding dual variables into a Lagrange dual bound to check for global optimality.

To check if a given solution is indeed a global solution, we exploit the fact that \eqref{eq: acots-lr} and its convex counterpart (with $X\succeq 0$) share the same Lagrange dual~\cite{richard-nns}. We derive the dual bound by dualizing the convex relaxation ($X=\sum\nolimits^{r}_{i=1} \rightarrow X\succeq0$) of \eqref{eq: acots-lr} and by adding a trace constraint on moment matrix $X$: ${\rm tr}(X)\le\rho$. We choose $\rho$ such that, at optimality, the trace constraint will not be binding~\cite{richard-nns}; it is added to the formulation to allow us to construct a guaranteed lower bound on the optimal solution. Defining the penalized cost matrix as
\begin{align}
F(\lambda,\mu,s)\triangleq C+\sum^{n}_{i=0}\lambda_{i}A_{i}-\sum^{m}_{i=1}\mu_{i}B_{i}-\sum_{i\in {\mathcal G}}y_{i}K_{i},
\end{align}
we construct the Lagrange dual via
\begin{align*}
\max_{\lambda,\mu\ge0,s\in\mathcal{K}^{*}}\min_{X\succeq0,{\rm tr}(X)\le\rho,t}t(1\!-\!s_1)+\left\langle F_{\lambda,\mu,s},X\right\rangle -\lambda_{0}-\tfrac{s_2}{2},
\end{align*}
where $\mathcal{K}^{*}$ is the dual RSOC. We have added the trace constraint to invoke the dual norm on the inner minimization: the trace-constrained minimization over a PSD matrix has a closed-form solution given by
\begin{align}
\min_{X\succeq0,{\rm tr}(X)\le\rho}\left\langle M,X\right\rangle =\rho\cdot \min(0,\lambda_{{\rm min}}\{M\}),
\end{align}
where $\lambda_{{\rm min}}\{M\}$ is the smallest eigenvalue of matrix $M$. This property was exploited, via bundle methods, in~\cite{bundle-bounds} to partially close the duality gap in SDP-based OPF problems. Applying this to the inner minimization of the Lagrange dual yields
\[
\max_{\lambda,\mu\ge0,s\in\mathcal{K}^{*}}\min_{t}\,t(1-s_{1})+\rho\cdot\min(0,\lambda_{{\rm min}}\{F_{\lambda,\mu,s}\})-\lambda_{0}-\tfrac{s_2}{2}.
\]
At optimality, however, $s_1=1$~\cite{activate}. Applying this, the inner minimization vanishes, and we are left with the bound
\begin{align}\label{eq: dual_bound}
d(\lambda,\mu,s)\triangleq\rho\cdot\min(0,\lambda_{{\rm min}}\{F(\lambda,\mu,s)\})-\lambda_{0}-\tfrac{s_2}{2}.
\end{align}
\begin{remark}
Set $s_1=1$. For any $\mu\ge 0$, $2\cdot s_2 \ge y^Ty$, and $\lambda$, \eqref{eq: dual_bound} will lower bound the solution to the low-rank SDP \eqref{eq: acots-lr}.
\end{remark}
\begin{remark}
By weak duality, if the dual bound $d(\lambda,\mu,s)$ of \eqref{eq: dual_bound} and the primal objective $c_p$ of \eqref{eq: acots-lr} match, then the associated low-rank solution must be a global solution.
\end{remark}
At global optimality, $F(\lambda,\mu,s)\succeq 0$; this exactly corresponds with SDP dual feasibility (i.e., the slack matrix must be PSD). Positive semidefiniteness directly implies $\lambda_{{\rm min}}\{F(\lambda,\mu,s)\}\ge0$, meaning the value of $\rho$\footnote{Since the trace of the moment matrix is the sum of nonnegative monomial terms involving voltages and binaries, it is straightforward to compute a reasonably tight value for $\rho$~\cite{bundle-bounds}.} will not affect the bound produced by \eqref{eq: dual_bound} if the candidate dual solution is in fact a global solution. If the candidate solution is not global, and $\lambda_{{\rm min}}\{F(\lambda,\mu,s)\}$ is significantly negative, this indicates that either the low-rank SDP solver got ``stuck" in a local minimum solution, or the low-rank search space was too small (i.e., $r$ must be increased).



\subsection{Global solutions to the AC-OTS problem}
We emphatically note that zero duality gap (i.e., $d(\lambda,\mu,s)\approx c_p$) does \textit{not} imply a globally optimal solution to the original, nonconvex AC-OTS problem; it simply means the low-rank SDP solution, which was generated by a nonconvex solver, is also the global solution to the convex relaxation. If we add a feasibility test, then we can claim AC-OTS optimality.

\begin{theorem}
    If a low-rank SDP solution for \eqref{eq: acots-lr} has zero duality gap $d(\lambda,\mu,s)\approx c_p$, and the solution matrix $X$ admits a solution that is feasible in the original AC-OTS problem \eqref{eq: ac-ots}, then the solution is the globally optimal solution. 
    \begin{proof}
    No superior solution can exist.
    \end{proof}
\end{theorem}

In the previous theorem, it can seem tempting to replace the ``primal feasibility" requirement with a rank-1 (or rank-2) condition on $X$, since this is a common condition in the OPF literature~\cite{molzahn2019survey}. However, \textit{underconstrained} portions of the lifted moment matrix need not fall into the rank-1 (or rank-2) solution basis, even if lower-degree portions of the matrix do admit a primal feasible solution. Furthermore, in our testing, we have seen the nonconvex solver numerically struggle to find ultra low-rank solutions, since the lifting routine can massively \textit{overconstrain} portions of the low-rank search space. 

\section{Case Study Test Results}
This section presents anecdotal test results to demonstrate the technical validity of the proposed approach. We test the proposed methodology on a 3-bus test case (\btt{pglib\_opf\_case3\_lmbd}) from PGLib~\cite{pglib}. The test case is slightly modified in two ways: ($i$) we impose branch current limits, rather than apparent power flow limits, and ($ii$) the line flow limit on branch 2 is lowered to 0.4 pu, to ensure the optimal switching decision turns a line off. To benchmark our approach, we formulated the nonconvex MIQCQP (81 equations total) and solved it to global optimality with Gurobi v13. This section does not necessarily demonstrate scalability; rather, it demonstrates that the proposed low-rank and lifted SDP approach can indeed find guaranteed global solutions to a nonconvex problem via a local solver.

\subsection{Lifting Results}
Table \ref{tab: lift} demonstrates the effectiveness of the lifting strategy. For demonstration purposes, we solved these lifted SDP problems with a mature conic solver (Mosek v11). We started by solving the vanilla SDP relaxation of \eqref{eq: acots-relaxed}. We then took 6 trips around the loop of Fig.~\ref{fig:sdp-lifting}, sequentially strengthening the relaxation via RLT cuts. By lift \#6, we arrived at a globally optimal solution with exact binary values and a rank-1 voltage profile. To the author's knowledge, this is the first time a conic solver has been used to find a global, feasible solution to a nonconvex AC switching problem in the power system literature. Notably, by the $6^{\rm th}$ lift, the formulation had attracted a huge number of constraints (2287 total), the vast majority of which were inactive. We tried removing almost 75\% of these constraints (based on dual variable magnitude), and we were able to solve for the same optimal solution.

\begin{table*}[t]
  \centering
  \caption{Successive lifts of the SDP relaxation, via Mosek, compared with the exact solution from Gurobi.}\label{tab: lift}
  \label{tab:sdp-lifts}
  \begingroup
  \setlength{\aboverulesep}{0pt}   
  \setlength{\belowrulesep}{0pt}
  \renewcommand{\arraystretch}{1.25}
  \setlength{\tabcolsep}{6pt}
  \begin{tabular}{c c c c c c c
                  >{\columncolor{liftgreen}}c
                  >{\columncolor{solvergray}}c}
    \toprule
    \toprule
    & \textbf{Vanilla SDP} & \textbf{Lift \#1} & \textbf{Lift \#2} & \textbf{Lift \#3} & \textbf{Lift \#4} & \textbf{Lift \#5} & \textbf{Lift \#6} & \textbf{Gurobi (Benchmark)} \\
    \midrule
    \textbf{Objective Value}      & 5958.2 & 6046.9 & 6132.0 & 6271.0 & 6317.4 & 6332.1
                   & \textbf{6344.6} & \textbf{6344.6} \\
    \textbf{\# of Constraints} & 141 & 274 & 1232 & 1783 & 2272 & 2281 & 2287 & 81 \\
    \bottomrule
  \end{tabular}
  \endgroup
\end{table*}

\begin{table*}[t]
  \centering
  \caption{Low-rank SDP solutions, via Knitro, for various rank $r$ search spaces.}
  \label{tab:lowrank}
  \begingroup
  \setlength{\aboverulesep}{0pt}   
  \setlength{\belowrulesep}{0pt}
  \renewcommand{\arraystretch}{1.25}
  \setlength{\tabcolsep}{6pt}
  \begin{tabular}{c >{\columncolor{liftred}}c >{\columncolor{liftred}}c >{\columncolor{liftred}}c >{\columncolor{liftred}}c >{\columncolor{liftyellow}}c >{\columncolor{liftyellow}}c
                  >{\columncolor{liftgreen}}c
                  >{\columncolor{liftgreen}}c
                  >{\columncolor{liftgreen}}c}
    \toprule
    \toprule
    & \textbf{Rank 1} & \textbf{...} & \textbf{Rank 6} & \textbf{Rank 7} & \textbf{Rank 8} & \textbf{Rank 9} & \textbf{Rank 10} & \textbf{Rank 11} & \textbf{Rank 12} \\
    \midrule
    \textbf{Primal Objective}      & infeas & ... & infeas & 6345.1 & 6344.6 & 6344.6 & 6344.6 & 6344.6 & 6344.6 \\
    \textbf{Termination Status}    & infeas & ... & infeas & time limit & local solve & local solve & local solve & local solve & local solve\\
    $\boldsymbol{\lambda_{\rm min}\{F(\lambda,\mu,s)\}}$ & - & ... & - & -80.1e3 & -4.49 & -1.35 & -2e-4 & -4e-4 & -2e-4\\
    $\boldsymbol{-\lambda_0-}\tfrac{\boldsymbol{1}}{\boldsymbol{2}}\boldsymbol{s_2}$ (***) & - & ... & - & 6345.1 & 6344.6 & 6344.6 & 6344.6 & 6344.6 & 6344.6 \\
    \bottomrule
  \end{tabular}
  \endgroup
\end{table*}

\subsection{Low-Rank Solution Results}
We then re-solved the lifted model from lift \#6 (with the majority of the inactive constraints removed) using Knitro~\cite{byrd2006knitro} as a low-rank SDP solver. We used relaxed optimizer configuration parameters from~\cite{richard-nns}. Results are summarized in Table~\ref{tab:lowrank}: for a sufficiently low-rank search space, Knitro struggles to find a solution. When the search space is expanded to $r=8$ and $r=9$, Knitro can find a solution, but the smallest eigenvalue of the slack matrix $F$ is still far from 0. For $r\ge 10$, the magnitude of the negative slack matrix eigenvalues drops by orders of magnitude, indicating the existence of a certified global solution.

One notable caveat: the final row of Table~\ref{tab:lowrank} presents part of the dual bound from \eqref{eq: dual_bound}; however, the bound in \eqref{eq: dual_bound} is built on the assumption that $s_1=1$, as was proven in prior work~\cite{activate}. For the conic solver (Mosek), this held; however, for the low-rank solver, $s_1$ would quibble around 1. To compute the presented bound (final table row), we had to add the correction factor $+t(1-s_1)$. Future work will peg $s_1$ at 1 to prevent this quibbling.

\section{Conclusion}

This paper demonstrated that a local, low-rank SDP solver (Knitro) can provide a guaranteed globally optimal and feasible solution to a nonconvex, mixed integer polynomial optimization problem. We successfully solved a toy AC-OTS problem containing 81 constraints (in the original formulation). There are a number of directions for future work, including the inclusion of localizing matrices in the tightening routine, automated RLT cut selection, and the demonstration of scalable results.

\section{Acknowledgments}
We acknowledge helpful discussions about this and related work with Saba Rafiei, Dan Molzahn, and Amrit Pandey.

\section{AI Disclaimer}
AI was not used in the writing of this paper. Claude was used to develop several subroutine functions for SDP lifting.

\appendices

{\section{Squared Current Formulation}\label{AppA}}
Complex currents are related via
\begin{align}
\left[\begin{array}{c}
i_{ij}\\
i_{ji}
\end{array}\right] & = \left[\begin{array}{c}
f^{T}\\
t^{T}
\end{array}\right] v \triangleq \left[\begin{array}{cc}
\frac{y_{l}+y_{s}}{\tau^{2}} & -\frac{y_{l}}{\tau e^{-j\theta}}\\
-\frac{y_{l}}{\tau e^{j\theta}} & y_{l}+y_{s}
\end{array}\right]\left[\begin{array}{c}
v_{i}\\
v_{j}
\end{array}\right].
\end{align}
``From" side square current \eqref{eq:iij} may be computed via
\begin{subequations}
\begin{align}
I^{2}_{ij} & =i_{ij}i^{*}_{ij}=\left(f^{T}v\right)\left(f^{T}v\right)^{*}\\
 & =v^{T}ff^{H}v^{*}={\rm tr}\left(\left(v^{*}v^{T}\right)\left(ff^{H}\right)\right)\\
 & =f_{1}u_{1}+f_{2}u_{3}+f_{3}u_{2}+f_{4}u_{4}\label{eq: fu}
\end{align}
\end{subequations}
where we have defined
\begin{subequations}
\begin{align}
\left[\begin{array}{cc}
f_{1} & f_{2}\\
f_{3} & f_{4}
\end{array}\right] & \triangleq ff^{H},\\
\left[\begin{array}{cc}
u_{1} & u_{2}\\
u_{3} & u_{4}
\end{array}\right] & \triangleq\left[\begin{array}{cc}
v^{*}_{i}v_{i} & v^{*}_{i}v_{j}\\
v^{*}_{j}v_{i} & v^{*}_{j}v_{j}
\end{array}\right]=v^{*}v^{T}.
\end{align}
\end{subequations}
The trace of the product of Hermitian matrices is real, so all imaginary values cancel out in \eqref{eq: fu}. ``To" side square current \eqref{eq:iji} can be computed analogously.

{\section{Big-M Tightening}\label{AppB}}
We use multiple rounds of optimization-based bound tightening (OBBT)~\cite{obbt} to sequentially maximize and minimize eight big-M bounds (upper and lower active and reactive flow in both directions) per line:
\begin{align}
\left.\begin{array}{c}
\overline{M}^{s}_{k}\\
\underline{M}^{s}_{k}
\end{array}\right\} ,\;s\in\{p,q\},\;k\in\{ij,ji\}.
\end{align}
To optimize these bounds, we took the SDP formulation, added the big-M linearization \eqref{eq: bigm2} for all line flows (power and current), relaxed binaries between 0 and 1, and added a local cost cut $c^{*}\ge\sum_{i\in\mathcal{G}}\left\langle K_{i},X\right\rangle ^{2}+\left\langle C,X\right\rangle$, where $c^{*}$ was found by computing a local OPF solution with all lines fixed ``on". Next, the SDP was relaxed into an RSOC (i.e., the standard $w_i w_j \ge c^2_{ij} + s^2_{ij}$ Jabr relaxation~\cite{activate} was applied). The resulting RSOCs were sequentially solved via Gurobi until subsequent tightenings didn't meaningfully move the bounds. The SDP was relaxed into an RSOC because solving SDP-based OBBT is generally too slow (there are 8 bounds per line, and many rounds of bound tightening).


\bibliographystyle{IEEEtran}
\bibliography{references}

@INPROCEEDINGS{mixed-sdp-rsoc,
  author={Molzahn, Daniel K. and Hiskens, Ian A.},
  booktitle={2015 IEEE Eindhoven PowerTech}, 
  title={Mixed SDP/SOCP moment relaxations of the optimal power flow problem}, 
  year={2015},
  volume={},
  number={},
  pages={1-6},
  doi={10.1109/PTC.2015.7232429}}

@INPROCEEDINGS{det-cuts,
  author={Hijazi, Hassan and Coffrin, Carleton and Van Hentenryck, Pascal},
  booktitle={2016 Power Systems Computation Conference (PSCC)}, 
  title={Polynomial SDP cuts for Optimal Power Flow}, 
  year={2016},
  volume={},
  number={},
  pages={1-7},
  doi={10.1109/PSCC.2016.7540908}}

@INPROCEEDINGS{lin-acopf,
  author={Bienstock, Daniel and Villagra, Matías},
  booktitle={2024 IEEE 63rd Conference on Decision and Control (CDC)}, 
  title={Accurate and Warm-Startable Linear Cutting-Plane Relaxations for ACOPF}, 
  year={2024},
  volume={},
  number={},
  pages={5024-5031},
  doi={10.1109/CDC56724.2024.10886304}}

@techreport{ferc,
  author      = {Gisin, Boris and Gu, Qun and David, Jim},
  title       = {Software Solutions for Increasing Market and Planning Efficiency},
  institution = {Federal Energy Regulatory Commission (FERC)},
  year        = {2017},
  type        = {Presentation for Technical Conference on Increasing Market and Planning Efficiency through Improved Software},
  address     = {Washington, DC},
  number      = {Docket No. AD10-12-008, Paper M3-3},
  url         = {https://www.ferc.gov/sites/default/files/2020-08/M3-3_Gisin.pdf}
}

@article{Holzer2025,
  author  = {Holzer, Jesse T. and Elbert, Stephen and Mittelmann, Hans and O’Neill, Richard and Oh, HyungSeon},
  title   = {Go competition challenge 3: problem, solvers, and solution analysis},
  journal = {Energy Systems},
  year    = {2025},
  month   = {01},
  day     = {31},
  issn    = {1868-3975},
  doi     = {10.1007/s12667-024-00708-1},
  url     = {https://doi.org/10.1007/s12667-024-00708-1}
}

@ARTICLE{go-kyri,
  author={Sharadga, Hussein and Mohammadi, Javad and Crozier, Constance and Baker, Kyri},
  journal={IEEE Transactions on Industry Applications}, 
  title={Scalable Solutions for Security-Constrained Optimal Power Flow With Multiple Time Steps}, 
  year={2025},
  volume={61},
  number={3},
  pages={4812-4821},
  doi={10.1109/TIA.2025.3532927}}

@article{spot-markets,
title = {Getting prices right on electricity spot markets: On the economic impact of advanced power flow models},
journal = {Energy Economics},
volume = {126},
pages = {106968},
year = {2023},
issn = {0140-9883},
doi = {https://doi.org/10.1016/j.eneco.2023.106968},
url = {https://www.sciencedirect.com/science/article/pii/S0140988323004668},
author = {Martin Bichler and Johannes Knörr}}

@article{lass,
author = {Lasserre, Jean B.},
title = {Global Optimization with Polynomials and the Problem of Moments},
journal = {SIAM Journal on Optimization},
volume = {11},
number = {3},
pages = {796-817},
year = {2001},
doi = {10.1137/S1052623400366802},
URL = {https://doi.org/10.1137/S1052623400366802},
eprint = {https://doi.org/10.1137/S1052623400366802}}

@article{Nie2014,
  author  = {Nie, Jiawang},
  title   = {Optimality conditions and finite convergence of Lasserre’s hierarchy},
  journal = {Mathematical Programming},
  year    = {2014},
  month   = {08},
  day     = {01},
  volume  = {146},
  number  = {1},
  pages   = {97--121},
  issn    = {1436-4646},
  doi     = {10.1007/s10107-013-0680-x},
  url     = {https://doi.org/10.1007/s10107-013-0680-x}
}

@INPROCEEDINGS{dan-moment,
  author={Molzahn, Daniel K. and Hiskens, Ian A.},
  booktitle={2014 Power Systems Computation Conference}, 
  title={Moment-based relaxation of the optimal power flow problem}, 
  year={2014},
  volume={},
  number={},
  pages={1-7},
  doi={10.1109/PSCC.2014.7038397}}

@article{josz,
author = {Josz, C{\'e}dric and Molzahn, Daniel K.},
title = {Lasserre Hierarchy for Large Scale Polynomial Optimization in Real and Complex Variables},
journal = {SIAM Journal on Optimization},
volume = {28},
number = {2},
pages = {1017-1048},
year = {2018},
doi = {10.1137/15M1034386},
URL = {https://doi.org/10.1137/15M1034386},
eprint = {https://doi.org/10.1137/15M1034386}}

@misc{verify-sdp,
      title={Verifying Global Optimality of Candidate Solutions to Polynomial Optimization Problems using a Determinant Relaxation Hierarchy}, 
      author={Sikun Xu and Ruoyi Ma and Daniel K. Molzahn and Hassan Hijazi and Cédric Josz},
      year={2021},
      eprint={2101.00621},
      archivePrefix={arXiv},
      primaryClass={math.OC},
      url={https://arxiv.org/abs/2101.00621}, 
}

@article{prove-opt,
title = {Proving global optimality of ACOPF solutions},
journal = {Electric Power Systems Research},
volume = {189},
pages = {106688},
year = {2020},
issn = {0378-7796},
doi = {https://doi.org/10.1016/j.epsr.2020.106688},
url = {https://www.sciencedirect.com/science/article/pii/S0378779620304910},
author = {S. Gopinath and H.L. Hijazi and T. Weisser and H. Nagarajan and M. Yetkin and K. Sundar and R.W. Bent}}

@article{bundle-bounds,
title = {Certified and accurate SDP bounds for the ACOPF problem},
journal = {Electric Power Systems Research},
volume = {212},
pages = {108278},
year = {2022},
issn = {0378-7796},
doi = {https://doi.org/10.1016/j.epsr.2022.108278},
url = {https://www.sciencedirect.com/science/article/pii/S0378779622004692},
author = {Antoine Oustry and Claudia D’Ambrosio and Leo Liberti and Manuel Ruiz}}

@article{Burer2003,
  author  = {Burer, Samuel and Monteiro, Renato D. C.},
  title   = {A nonlinear programming algorithm for solving semidefinite programs via low-rank factorization},
  journal = {Mathematical Programming},
  year    = {2003},
  month   = {02},
  day     = {01},
  volume  = {95},
  number  = {2},
  pages   = {329--357},
  issn    = {1436-4646},
  doi     = {10.1007/s10107-002-0352-8},
  url     = {https://doi.org/10.1007/s10107-002-0352-8}
}

@misc{Burer2023TwoDecades,
  author       = {Burer, Sam},
  title        = {Two Decades of Low-Rank Optimization},
  year         = {2023},
  month        = {09},
  day          = {09},
  howpublished = {YouTube},
  url          = {https://www.youtube.com/watch?v=wSauUgRIQDg}
}

@inproceedings{no-spurious,
author = {Ge, Rong and Jin, Chi and Zheng, Yi},
title = {No spurious local minima in nonconvex low rank problems: a unified geometric analysis},
year = {2017},
publisher = {JMLR.org},
pages = {1233–1242},
numpages = {10},
location = {Sydney, NSW, Australia},
series = {ICML'17}}

@misc{lorads,
      title={A Low-Rank ADMM Splitting Approach for Semidefinite Programming}, 
      author={Qiushi Han and Chenxi Li and Zhenwei Lin and Caihua Chen and Qi Deng and Dongdong Ge and Huikang Liu and Yinyu Ye},
      year={2024},
      eprint={2403.09133},
      archivePrefix={arXiv},
      primaryClass={math.OC},
      url={https://arxiv.org/abs/2403.09133}, 
}

@misc{culorads,
      title={Accelerating Low-Rank Factorization-Based Semidefinite Programming Algorithms on GPU}, 
      author={Qiushi Han and Zhenwei Lin and Hanwen Liu and Caihua Chen and Qi Deng and Dongdong Ge and Yinyu Ye},
      year={2024},
      eprint={2407.15049},
      archivePrefix={arXiv},
      primaryClass={math.OC},
      url={https://arxiv.org/abs/2407.15049}, 
}

@INPROCEEDINGS{lropf1,
  author={Huang, Shengquan and Bai, Xiaoqing and Shang, Qinghua and Wang, Rui},
  booktitle={2023 3rd Power System and Green Energy Conference (PSGEC)}, 
  title={A Low-Rank Algorithm Based on Riemannian Optimization for Optimal Power Flow Problem}, 
  year={2023},
  volume={},
  number={},
  pages={388-392},
  doi={10.1109/PSGEC58411.2023.10255934}}

@article{lropf2,
author = {Jakub Mareček and Martin Takáč},
title = {A low-rank coordinate-descent algorithm for semidefinite programming relaxations of optimal power flow},
journal = {Optimization Methods and Software},
volume = {32},
number = {4},
pages = {849--871},
year = {2017},
publisher = {Taylor \& Francis},
doi = {10.1080/10556788.2017.1288729},
URL = {https://doi.org/10.1080/10556788.2017.1288729},
eprint = {https://doi.org/10.1080/10556788.2017.1288729}}

@misc{activate,
      title={Activate the Dual Cones: A Tight Reformulation of Conic ACOPF Constraints}, 
      author={Saba Rafiei and Samuel Chevalier},
      year={2026},
      eprint={2603.20411},
      archivePrefix={arXiv},
      primaryClass={eess.SY},
      url={https://arxiv.org/abs/2603.20411}, 
}

@ARTICLE{boxalg,
  author={Kundert, K.S. and Sangiovanni-Vincentelli, A.},
  journal={IEEE Transactions on Computer-Aided Design of Integrated Circuits and Systems}, 
  title={Simulation of Nonlinear Circuits in the Frequency Domain}, 
  year={1986},
  volume={5},
  number={4},
  pages={521-535},
  doi={10.1109/TCAD.1986.1270223}}

@manual{mosek_modeling_cookbook,
  title     = {MOSEK Modeling Cookbook},
  author    = {{MOSEK ApS}},
  edition   = {Latest},
  year      = {2026},
  url       = {https://docs.mosek.com/modeling-cookbook/index.html},}

@inproceedings{richard-nns,
  title={Tight certification of adversarially trained neural networks via nonconvex low-rank semidefinite relaxations},
  author={Chiu, Hong-Ming and Zhang, Richard Y},
  booktitle={International Conference on Machine Learning},
  pages={5631--5660},
  year={2023},
  organization={PMLR}
}

@article{molzahn2019survey,
  title={A survey of relaxations and approximations of the power flow equations},
  author={Molzahn, Daniel K and Hiskens, Ian A},
  journal={Foundations and Trends in Electric Energy Systems},
  volume={4},
  number={1-2},
  pages={1--221},
  year={2019},
  publisher={Emerald Publishing Limited}
}

@misc{pglib,
      title={The Power Grid Library for Benchmarking AC Optimal Power Flow Algorithms}, 
      author={Sogol Babaeinejadsarookolaee and Adam Birchfield and Richard D. Christie and Carleton Coffrin and Christopher DeMarco and Ruisheng Diao and Michael Ferris and Stephane Fliscounakis and Scott Greene and Renke Huang and Cedric Josz and Roman Korab and Bernard Lesieutre and Jean Maeght and Terrence W. K. Mak and Daniel K. Molzahn and Thomas J. Overbye and Patrick Panciatici and Byungkwon Park and Jonathan Snodgrass and Ahmad Tbaileh and Pascal Van Hentenryck and Ray Zimmerman},
      year={2021},
      eprint={1908.02788},
      archivePrefix={arXiv},
      primaryClass={math.OC},
      url={https://arxiv.org/abs/1908.02788}, 
}

@incollection{byrd2006knitro,
  author    = {Byrd, Richard H. and Nocedal, Jorge and Waltz, Richard A.},
  title     = {Knitro: An Integrated Package for Nonlinear Optimization},
  booktitle = {Large-Scale Nonlinear Optimization},
  pages     = {35--59},
  year      = {2006},
  publisher = {Springer}
}

@article{go3Sam,
title = {A parallelized, Adam-based solver for reserve and security constrained AC unit commitment},
journal = {Electric Power Systems Research},
volume = {235},
pages = {110685},
year = {2024},
issn = {0378-7796},
doi = {https://doi.org/10.1016/j.epsr.2024.110685},
url = {https://www.sciencedirect.com/science/article/pii/S0378779624005716},
author = {Samuel Chevalier},
}

@misc{gurobi,
  author = {{Gurobi Optimization, LLC}},
  title = {{Gurobi Optimizer Reference Manual}},
  year = 2026,
  url = "https://www.gurobi.com"
}

@INPROCEEDINGS{obbt,
  author={Sundar, Kaarthik and Nagarajan, Harsha and Misra, Sidhant and Lu, Mowen and Coffrin, Carleton and Bent, Russell},
  booktitle={2023 62nd IEEE Conference on Decision and Control (CDC)}, 
  title={Optimization-Based Bound Tightening Using a Strengthened QC-Relaxation of the Optimal Power Flow Problem}, 
  year={2023},
  volume={},
  number={},
  pages={4598-4605},
  doi={10.1109/CDC49753.2023.10384116}}

\end{document}